# *Software engineering for mobile applications, a survey on challenges and solutions*

*Shehab Eldeen Ayman Mounir*
*Shehab.ayman@bue.edu.eg*
*The British University in Egypt*

**Abstract**

Mobile app development has become the front line in software engineering. With the recent years many smartphone platforms have grew including but not limited to webOS, blackberry os, Tizen, android, and iOS. The coexistence of these platforms results in a challenging situation where apps must be developed and maintained to the same level. The mobile app development scene has recently seen a noticeable rise in the number of applications that adapt web elements like HTML5 to produce native like applications that are essentially web views wrapped into containers to appear as any normal application. This means that the application behavior can vary drastically from one user to another meaning that the app behavior can be changed drastically. Therefore, application developers rely on an agile or an ad-hoc approach to development that is mostly autonomous. In this paper, we describe the current state of the art of context awareness in mobile application development.

**introduction**

No one can deny that the astronomical rise in smartphone popularity in the last decade has reshaped the software engineering industry. With billions in app downloads, mobile app development has become the front line in software engineering. Through the recent years many smartphone platforms have grew include but not limited to webOS, blackberry os, Tizen, android and iOS. With IOS and android taking most of the share of the smartphone platform. some of the common challenges of building a multifunctional software has moved from the desktop to mobile but developing software for smartphones has come with its own set of new challenges that needs to be overcome.

Recent estimations done in 2017 state that the apple app store contains more than 2.2 million apps while the google play store contains 2.8 million and they are in a constant growth. Even though both windows app store and blackberry world are both discontinued, they contained 600 thousand and 200 thousand apps respectively [1].

Developing for these mobile platforms is not as flexible as developing for the desktop. The programming languages, tools, frameworks and APIs required for the development process are platform specific, for instance, native android applications are built using the java programming language on the android studio integrated development environment, whereas apple's IOS applications are developed using the swift language with XCode tool [2]. The coexistence of these platforms results in a challenging situation where apps have to be developed and maintained to the same level.

Traditional software engineering approaches have been somewhat phased out in the context of mobile development as these approaches cannot be directly applied in the mobile context. Mobile graphical user interfaces are considered as a whole new paradigm for human computer interaction study. Mobile phones provide a whole new way of interaction that is considered untraditional comparing it the computer situation which involves sitting down and focusing attention and resources towards the interaction. And even the interaction method has drastically changed from the typical mouse and keyboard to touch screens and gestures. And the way of interaction seems to be evolving constantly as voice commands and augment reality/mixed reality are becoming more of a possibility [3]. Nowadays the variety of hardware and software platform have forced developers to make a group of different applications that on the outside might look the same but on the inside are completely different just to suite every platform [4].

**Challenges with mobile development**

1. Create a universal graphical user interface

Research conducted by Tarasewich and Gong suggest that at least four of Shneiderman's design principals -which will be looked at further- are applicable -without modification- to mobile phones. These include "enabling frequent users to use shortcuts", "designing dialogue to yield closure", "offering informative feedback", "supporting internal locus of control" [5]. As technologies tend to evolve, challenges tend to evolve as well that's why recently efforts have been into researching ways of streamlining application development regardless of the hardware or the software platform. Focusing on streamlining the process will free up a lot for resources which could then be directed towards making better user interfaces.

2. Reusing software across different platforms

Companies take the hard decision whether to focus on one or two platforms and enrich the user experience or go all out and support natively every single platform that is currently on the market compromising their software quality [6].

There is no other way of saying it but sometimes user interface cannot be unified across IOS and android not due to some technical limitations in any of the two platforms but due to the way each platform users have grown accustomed to certain design aspects. An application with an insert circular icon in the bottom right corner and three vertical dashes in the top left corner - that when clicked bring a charms menu- is easily identified by mobile developers as an android application. While applications that have their back/previous button in the top right corner tend to be IOS applications. Android applications tend not to have a back button as android users are accustomed to using the back button implemented in the operating system itself. While IOS users are accustomed to look for a back button in the application itself and when it is not present, a swipe to the right would function exactly like a back button.

Many software companies have separate teams for development where each time only focuses on one single platform. This means that if an app is to be developed for IOS and android -for example- the software engineering effort needed to build the app are doubled to provide the same functionalities on the two operating systems. The basic software engineering concepts of reusing and refactoring parts of software are not applicable here as they cannot be transferred from one software to another. This leads to a very limited coordination between development teams. It mostly relies on ad-hoc basis without any real effort in reducing the resources -especially time and cost- allocated for the project.

The mobile development scene has recently seen a noticeable rise in the number of applications that adapt web elements like HTML5 to produce native like applications which are essentially web views wrapped into containers to appear as any normal application. These kind of applications -which will be discussed further- do not have the rich capabilities of native apps as they cannot access the platform's APIs which limits their functionalities, but on the other side allow for the reuse of almost an entire application interface on more than one platform [7].

3. Context aware applications

Mobile devices are far in contrast to traditional stationary computer platforms. Mobile phones are highly customizable and adaptive to user's needs which means that they must constantly monitor the environment. This means that applications on mobile need to be constantly aware of time, weather, location, orientation, proximity, etc. [8] Mobile applications are now able to use all of this contextual awareness to make a very specific, very specialized experience that is very special to its own user meaning that the application behavior can vary drastically from one user to another. This idea of context awareness is no new feature, web applications have been providing similar kind of user tailored experience for years but it has never got to the same extent as mobile applications due to the sensors and capabilities of mobile phones. This kind of sensory overload has never been present or dealt with in traditional software engineering, it was always associated with robotics and smart objects which are dealt with using agent oriented software engineering [9]. The availability of these sensors has put their utilization in the forefront of application development. So, developers pour careful attention into analyzing application requirements and utilizing context awareness to much improve the quality of their application resulting in a better user experience.

Agent oriented software engineering (AOSE) is concerned with building software agents that are mostly autonomous. This approach provides all the necessary models, abstractions and "pure" software engineering approaches to build a contextually aware application of a multi-agent system (MAS) [10]. For simplification lets imagine that the multiagent system is a physical robot. Using its sensors, the robot must use all of its sensors to sense and understand its surrounding environment and react to it accordingly in order to achieve its goals.

4. Balancing agility and uncertainty

There is no denying that the existence of mobile applications has changed the way a development process is looked at. Gone are the times where almost all the application requirements are clearly specified and fully implemented on the first run. It is now completely normal to regularly receive updates for applications that add new features or focus on stability improvement and bug fixes. This means that the traditional waterfall model cannot be realistically applied in mobile development that's why application developers rely on an agile or an ad-hoc approach to development. The constant growth in demand for more context aware applications, tailored user experience, very heavy competition and low tolerance by users for unresponsive applications have derived more of a semi-formal approach development that does not necessarily go by the books. This approach has to be somehow integrated within the agile method. The very dynamic very user specific experience provided my mobile applications allows for scenarios and situations that may not be fully specified within the functional and nonfunctional requirements. If mobile applications were strictly to follow their functional / nonfunctional requirements, this will result in a less quality experience. This means that applications need to run continuously and autonomously adapt and modify the behavior and provide more functionality than strictly specified.

It is hard to imagine the idea of how applications can provide an improved service than originally specified within the design documents and how this constantly adaptive service can improve the quality and engagement with applications. A great example of adaptive service is user tailored ad experience. This has become an essential way of digital advertising where the products listed in the ad spots depending on the user interests. Gathering user interests has lately been a relatively easy task. For example, google ad services track each user's search history to determine what are this specific user's interest and then display ads of objects that the user may find tempting to buy. Another great example is Facebook application, it uses location services and profile analysis to recommend friends. And this context awareness feature keeps updating frequently. For example, if a person left his workplace, university or even the country, the application will keep providing relevant and new recommendations based on the updated location. Another area where Facebook has ventured into and has made a big impact is with the Facebook marketplace service on which people can buy and sell almost anything from used phones to used cars. The application here utilizes context awareness to link users with the nearest buyers and promote products to users who are interested in as it will most likely end up in a successful buy / sell operation.

**Cross platform development**

An ongoing challenge in mobile development especially for startups and relatively small companies is to choose the platforms that they are going to develop for. Companies tend to focus their development on IOS and android to cover the largest amount of smartphone users possible. Having the application accessible to the most possible number of users brings more profit and makes more of an impact on the market. So, it is time and money consuming for companies to hire skilled developers that can develop applications up to the same standard on each platform.

**Challenges with creating cross platform application development**

1. Creating a universal graphical user interface

Regardless from the actual design itself, every mobile platform provides its own way for developers to address user interface requirements and manipulating in to developer's needs. A good start would be to cope with the variety screen sizes and resolutions of different smartphones and tablets. Android for instance is very flexible with this aspect as it gives developers more flexibility when dealing with screens with different sizes and resolutions. Unlike Apple which seems to be very rigid in this aspect. IOS applications are restricted in size and resolution based on the specific iPhone/iPad models that are targeted. Thus, a unified user interface design is a bit of a challenge for developers to implement across different platforms.

a survey published in the international journal of advanced computer science and applications interviewed a diverse group of smartphone application developers with different years of experience to see how they approach UI design. It was noticed in the survey that IOS developers tend to follow Apple's own UI guidelines which apply constraints on sizes and resolutions but also give some flexibility in how to approach this design whether to rely on apple's own XIB with storyboards or MVC. On the other hand, android developers seemed to follow Google's own guidelines on material design as it provides better user experience [11].

2. Issues within a single platform

Within a single platform, internal challenges may arise that could make development even more difficult. Android is a perfect example for that. A wide variety of android operating system flavors co-exist in the market. Samsung, Xaiomi, Oppo and Huawei each have their own skin on top of the android operating system which sometimes leads to incompatibilities with certain parts of applications. A new dilemma has unfolded in early 2020 with the trade wars between the united states of America and china has affected software development. Due to recent developments, Chinese telecommunication company Huawei can no longer take advantage of American made software. Although Huawei would still use the android operating system, they get to lose out on Google's complimentary services like the Google play store and Google play services. This issue has caused a lot of problems for companies as now decision makers have to choose whether to integrate google play services which are arguably essential to develop an application that is up to the highest standards or to ditch the services in favor of making an app that can be published independent from Google's own play store. This is a tough decision to take as decision makers have to choose whether to integrate the services and appeal to north American and European users and lose a potential of a billion smartphone users in south east Asia or develop an app that is available in the Asian market and risk providing an app that is substandard in quality compared to competitors applications.

There exist some cross-platform frameworks that aim to ease the development process for making different versions of the same app to run on completely different platforms. The original goal of any cross-platform framework is to "write once compile more" but the success of a framework is not always guaranteed. One of most famous cross-platform frameworks is Xamarin. It is developed my Microsoft and it supports both IOS and Android. The programming language used to write for both platforms is C#. developers write the apps as if they are developing for windows phone devices. Using compatibility libraries, Xamarin then takes

Windows SDK API calls and translates them to their equivalent API calls for IOS and android. Another new framework that has become a hot topic is React-Native which is developed by Facebook and also aims to provide a cross-platform development environment which supports both IOS and android.

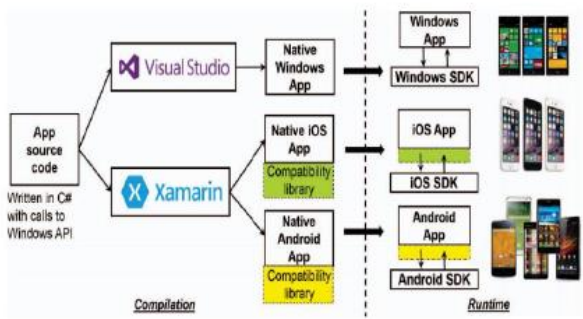

Figure 1 shows how source code is transformed to IOS and android code [12]

Developers often face a lot of challenges when trying to use these cross-platform frameworks. These frameworks often tend to lack proper support for their own libraries so when developers get stuck with an issue, they find it hard to get the proper community support. And some of the libraries themselves tend to have limited power when trying to access mobile-specific functionality and proper hardware utilization which leads to memory leakage issue and less CPU utilization.

It is hard to find good cross-platform developers and especially those needed to write to a specific framework and have a background in writing, integrating and testing in previous projects. Projects that involve great risks or constraints - whether time or money constraints- tend to stick to the more stable and secure native development approach and avoid cross-platform.

In the survey some of the developers stated that sticking to native applications makes for an easier development process as there is a better community support and it takes a lot of time and effort to make a cross-platform app imitate the look and feel of a native app. Some other developers said that they were open to learn new ways to develop applications but the environment they are in is not really helping. Companies with big projects tend to be constrained on time and money when it comes to delivering big projects, so it puts the project at risk if it was to be developed in a relatively new - new to the developers- environment. One developer said that a good developer if given the proper time and resources would not find it difficult to learn any of the new cross-platform frameworks. Another developer said that learning a framework for cross-platform development should not be hard at all as all of them share the same basic object-oriented programming concepts that lay the foundation for programming.

3. Changes in requirement

Whether it is a desktop application or a mobile application, if the client changes or modifies some requirements during the initial stage of development, these modifications can be made. If the requirements are constantly changing while enough time is given for development and proper testing then there is still no problem. But of requirements change in a late stage of the development life cycle then the associated cost of fixing them would be quite high. Frequent changes in development whether due to vague requirements from the first place or the sudden change in client needs is very costly because these changes have to be taken into consideration and sometimes the whole development process would need to be restarted from the beginning. Unclear changes waste time and money that's why agile processes have slots for coping with these kinds of occasions [13].

4. The importance of testing during development

Testing is a very important part of the development process; its aim is to find and correct the errors that are present in the application to give more effect. Testing in general aims to increase the quality of the application. But small organizations cannot really afford to allocate resources to testing. And sometimes small organizations cannot really afford to hire testers so developers double as testers as well [14]. And it is generally hard for a developer to find bugs in a code that he wrote himself. While organizations may not find the resources to test at all, other organizations test their applications only on emulators. Emulators by nature are very limited in features, they lack in terms of mobility, dynamic resolutions and aspect ratios and mobile oriented features like compass, gyroscope and GPS.

Through the survey it has been concluded that the ways of testing vary from one company to another. Some companies preferred automated testing while others preferred manual testing. Other companies follow load, alpha, beta and smoke testing.

One developer stated that their company follows a scrum methodology even though they understand that it is not a good practice for their relatively small apps. While another developer said that their company has a dedicated QA department which manually tests the apps on actual devices. In general, almost all the developers agreed on the idea that testing is an absolute necessity to deliver a bug free app.

5. Adapting old APIs to support new features

Another big dilemma that faces developers when trying to make an app -whether native or cross platform- is whether or not to utilize bleeding edge technologies. These technologies may be appealing to implement in the app but the problem always lies with supporting old software and old hardware because most of the app users will not have access to the latest resources so this has to be put in mind. A way of getting around the problem is using support libraries which help bring new features from new APIs to old APIs.

6. Maintenance

Another challenge in developing any kind of software whether for desktop or mobile Is to maintain the app and analyze crashes. Due to the almost infinite combination between hardware and software it is hard to replicate the crashes for the purpose of analysis and fixing.

**Research efforts towards better software engineering for mobile devices**

1. Designing special user interface for people with accessibility issues

According to a report published by the US Census, about 15 percent of people living in the united states suffer from at least one disability. This disability can be physical, sensory or other. Very few advancements in research has been conducted in this area with regards to human mobile interaction due to the limited studies that show communities' relations with their mobile phones [15].

There exist some general guidelines on how to modify existing applications to assist people with visual impairment. Those guidelines mostly suggest using "text to speech" services to assist with low vision / blind users. A text to speech service acts as a screen reader, it reads whatever is on the screen with a clear and loud voice which eases the developers work as no significant modification needs to be made to the design of an application to make it work. For example, Apple has built a "voice over" service inside IOS platform so it can work on almost all applications [16].

2. Software product line engineering

One of the development approaches that are geared towards bringing down development cost is software product line engineering. The aim of this approach is to determine and group applications that are similar in functionality and reuse existing code / logic across them. This approach can be considered the most successful approach yet for efficient development in the era of multiplatform applications [17].

The book "Software product line engineering: a family-based software engineering process" highlights two main phases for software product line engineering which are domain engineering phase and application engineering phase. Assuming that a software company has defined its product line, the domain phase then defines both the common and application specific requirements for the whole product line. Then comes the application phase where actual coding takes place, the common functionality logic is shared and used across the entire product line. The product line might not consist of entirely different applications it may be the same application but on different platforms. This approach really encourages developers to focus more on the common elements in the product line such as requirements and design. It helps developers understand the requirements regardless which specific platform the software is going to be written to. This approach also shifts the development process by putting requirement gathering and understand upfront rather than starting with design approaches which -in theory- should yield a faster development process.

3. Self-adaptive requirements

By nature, functional requirements tend to take all the attention and focus of developers and clients from non-functional requirements which are as critical as their counterparts [4]. Applications need to constantly "self-adapt" to provide users with the functionality they need. In some situations, applications need to self-adapt to provide users with reduced functionality to provide a better support for dynamism.

A good approach is to use already existing self-adaptive system requirement specification languages such as "RELAX" [18]. The language "Relax" was designed as a middleware language that aims to explicitly express the behavioral and environmental uncertainties that are come attached to self-adaptive systems. Relax takes a simple approach to dividing up requirements into those that must be satisfied completely (invariant) and those whose partial satisfaction / completion is not a necessity (variant). All the requirements are then expressed using a combination of natural language and fuzzy logic. The RELAX documentation then lists out the variant requirements and how the environment can affect them.

Integrating a self-adaptive specification language like RELAX into an existing agile software engineering approach would provide a better structure for requirement gathering and specification and would overall improve the quality of non-functional system requirements in the context of environment changes. Adapting this approach will allow developers to better consider and adapt their application's behavior in a non-optimal environment.

**Shneiderman's Eight Golden Rules to design better interfaces**

About Shneiderman's

Ben Shneiderman is an American computer science professor how specializes in human computer interaction in the University of Maryland. He made an infamous book about human computer interaction called "Designing the User Interface: Strategies for effective human-computer Interaction" in which he outlines eight golden principals for designing an interface. Four of these principles can be directly translated into application design while the other four -while still relevant- would need some modifications to translate to mobile app development [19].

1. Strive for consistency

Consistency in design relies on using same design elements through the entire application. Design elements are the icons, menu hierarchy, colors and actions. Users need to easily understand the flow of the application and how they can do a sequence of actions to achieve their goals. The experience gained from interaction with a page needs to be transferrable and applied to the other pages within the same application without needing to learn a new skill. Consistency in design helps with the feeling of familiarity with the software product.

A better way to implement a consistent design is not only to have consistency within all the pages of the application but also to have consistency with the design philosophy of the platform. One example that has been adapted by many android developers is the charms menu. It is very often now on android applications is to find on the top left corner 3 parallel horizontal dashes that when pressed bring up a sliding charms menu. This menu contains all the necessary options from settings to account management. The main reason android developers use the charms menu is because it is recommended by google as part of the "material design" interface. Utilizing material design elements means that users can translate their experience from one application to another making the interaction easier.

2. Enable frequent users to use shortcuts

An important demographic that must not be ignored when designing an application is the power users. Power users do not look or behave similar to casual users. For power users the entire phone is just a utility to achieve multiple goals and sometimes multiple goals at the same time. So, for power users, there has to exist shortcuts that would enable them to achieve their goals in a shorter time and with less steps.

A great implementation on shortcuts is Apple's built in shortcut app on IOS. The application enables users to automate almost any kind of tasks on IOS. These tasks can be combined in any order and be are contextually aware so they can be triggered by time or location. These tasks can range from morning chores to congratulating someone on a special occasion. A great example is to setup a morning routine that triggers every morning on an exact time. It would start with an alarm to wake up the user then inform him / her with set day's temperature and the probability of rain. It could go on to then read the important news of the day, turn on the coffee machine and order an uber for work. After the user leaves for work, the phone then proceeds to lock the house doors using the smart lock. This kind of automation was in recent years considered a future dream that has become a reality due to the advancements in mobile development.

3. Offer informative feedback

A proper engaging feedback should inform the user of what actions they made and also it should come in a proper timely manner. A good example for a good feedback is a processing status bar that would indicate to the user how long a certain process will take. A very bad example, is to have an application that produces error codes whenever it encounters errors. Error codes are not understood by users which makes the experience frustrating.

4. Design dialogue to yield closure

Communication is key, it is how humans understand each other and exchange information. The same principals should be applied when developing an application. For example, a form of reassurance has to be present after completing a transaction process. It would be better if the application could display a message confirming that the transaction has been completed successfully or if it has failed. Even if it has failed the application should display a message indicating that it had failed and explaining where exactly an error has happened.

5. Offer simple error handling

Developers should never expect that their applications are used by power users and experienced testers. An application should be fool-proof and even in a rare case of an application breaking error, the user should be guided by an easy step by step solution that he / she could follow.

6. Permit easy reversal actions

Continuing on with the idea of not expecting application users to be experts, there should always exist a way to reverse an action or a sequence of actions in the cases of accidental presses. The undo button is frequently used in photo editing applications where users might accidently apply the wrong filter on a picture.

7. Support internal locus of control

A good way of interaction between users and an application is to give users sense of control and that they always initiate the actions.

8. Reduce short term memory load

Application design interfaces should be simply organized with a proper layout and hierarchy. The design should encourage the user to rely on recognition not recall as it is much easier. Using the application should not feel like a puzzle or a brain exercise. An application should always give users visual or auditory clues and provide relevant information at all times.

**conclusion**

Mobile platforms are moving towards more and more fragmentation whether its due to the variety of software platforms or even to the different flavors within each platform. An application can have the same name, design, look and feel on two different platforms but still be treated as two different applications due to the nature of development for each platform. So, to ensure consistency, functionalities of each app have to be manually checked against similar versions of it on the other platforms. Creating a general-purpose reusable graphical user interface depends on balancing compromises whether to be consistent or to follow the design language of each platform. Testing is still considered as a huge challenge among the mobile app development industry. Unit testing is still not common among the mobile development community and the current testing structures are not mature enough. But not all the blame has to be put on developers as the current mobile testing tools look very powerless as they tend to lack the fundamental features of mobile phones like sensors, gesture navigation support and location support.